\documentclass[twocolumn,showpacs,prb]{revtex4}

\usepackage{graphicx}
\usepackage{amssymb}
\usepackage[tbtags]{amsmath}

\begin{document}

\title{Typical versus average helicity modulus in the three-dimensional
gauge glass:\\[1mm] Understanding the vortex glass phase}

\author{Helmut G.~Katzgraber}
\affiliation
{Theoretische Physik, ETH Z\"urich,
CH-8093 Z\"urich, Switzerland}

\author{D.~W\"urtz}
\affiliation
{Theoretische Physik, ETH Z\"urich,
CH-8093 Z\"urich, Switzerland}

\author{G.~Blatter}
\affiliation
{Theoretische Physik, ETH Z\"urich,
CH-8093 Z\"urich, Switzerland}

\date{\today}

\begin{abstract}

We numerically compute the helicity modulus of the three-dimensional
gauge glass by Monte Carlo simulations. Because the average free
energy is independent of a twist angle, it is expected that the average
helicity modulus, directly related to the superfluid density, vanishes
when simulations are performed with periodic boundary conditions. This
is not necessarily the case for the typical (median) value, which
is nonzero, because the distribution of the helicity modulus among
different disorder realizations is very asymmetric. We show that
the data for the helicity modulus distribution are well described
by a generalized extreme-value distribution with a nonzero location
parameter (most probable value). A finite-size scaling analysis of the
location parameter yields a critical temperature and critical exponents
in agreement with previous results. This suggests that the location
parameter is a good observable.  There have been conflicting claims as
to whether the superfluid density vanishes in the vortex glass phase,
with Fisher {\em et al.}~[Phys.~Rev.~B {\bf 43}, 130 (1991)] arguing
that it is finite and Korshunov [Phys.~Rev.~B {\bf 63}, 174514 (2001)]
predicting that it is zero. Because the gauge glass is commonly used
to describe the vortex glass in high-temperature superconductors,
we discuss this issue in light of our results on the gauge glass.

\end{abstract}

\maketitle

\section{Introduction}
\label{sec:introduction}

The gauge glass is often used to describe the extremely
disordered vortex glass phase in granular high-temperature
superconductors.\cite{blatter:94,brandt:95,nattermann:00,blatter:03}
In addition, it is the simplest possible model that possesses the
correct order parameter symmetry [$XY$ spins with a U(1) symmetry]
and therefore considerable numerical, as well as theoretical, work
has been based on it.  Nevertheless, the model has some limitations:
it is isotropic, whereas finite magnetic fields in the vortex
glass phase in high-temperature superconductors introduce a uniaxial
field anisotropy.\cite{aniso} Furthermore, the gauge glass ignores
transverse screening, an essential ingredient in superconductivity. In
fact, the inclusion of screening destroys the gauge glass phase and
the critical temperature drops to zero.\cite{wengel:96} In addition,
the disorder is introduced via random gauge fields, whereas a more
realistic model would probably introduce the disorder via random
bonds.\cite{spivak:91,kawamura:00}

\begin{figure}[!tbp]
\includegraphics[width=7.5cm]{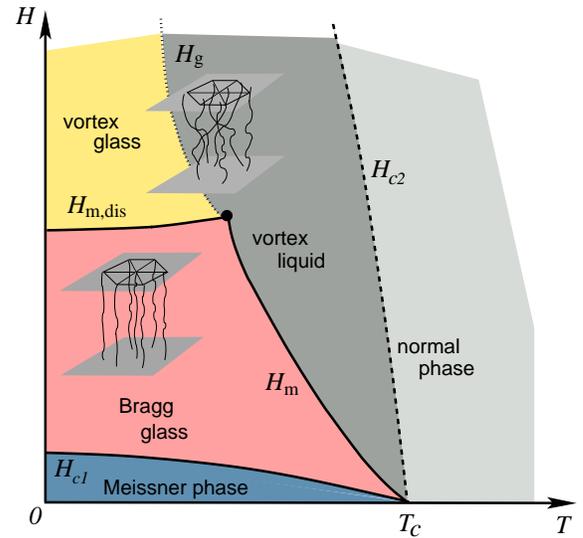}
                                                                                
\caption{(Color online)
Possible phase diagram of a disordered superconductor
(Ref.~\onlinecite{blatter:03}).  The vortex lines are pinned by
defects and the vortex solid found in a clean superconductor
is transformed into a glass. For low disorder, there are no
dislocations and one obtains a Bragg glass which exhibits a
superconducting response, whereas for high disorder and low
enough, temperatures a highly disordered vortex glass phase emerges
(Ref.~\onlinecite{giamarchi:94}). The vortex liquid remains liquid
after the inclusion of disorder, albeit viscous and separated by
a glass transition line $H_{\rm g}$ (dotted line) from the vortex
glass phase.  $H_{{\rm c}1}$ separates the Bragg glass phase from the
Meissner phase, whereas $H_{{\rm c}2}$ denotes the crossover from the
superconducting to the normal phase (Ref.~\onlinecite{gennes:89})
(mean-field calculation, dashed line).  The melting lines between
Bragg glass, vortex glass, and vortex liquid are denoted by $H_{\rm m}$
and are calculated taking into account fluctuations. (Figure adapted
from Ref.~\onlinecite{blatter:03}).
}
\label{fig:pd}
\end{figure}

In three space dimensions the (unscreened) gauge glass
undergoes a phase transition into a glassy phase at $T_{\rm c} =
0.47(3)$,\cite{olson:00} which has been estimated by a finite-size
scaling analysis of the currents in the system (first-order
derivative of the free energy with respect to an infinitesimal
twist to the boundaries).\cite{olson:00} Another central quantity to
study is the helicity modulus---directly related to the superfluid
density---which corresponds to the second-order derivative of the
free energy with respect to an infinitesimal twist to the boundaries.
Motivated by contradicting predictions as to whether the superfluid
density $\rho_{\rm s}$ is finite or not for a bulk vortex glass
system,\cite{fisher:91b,korshunov:01} we have studied in detail the
helicity modulus of the three-dimensional gauge glass model.  Because
the simulations are performed using periodic boundary conditions
in order to avoid finite-size effects due to small system sizes,
we expect that the average of the helicity modulus is zero. This
(unphysical) result follows from twists forced into the system due to
the used periodic boundary conditions.  The problem can be alleviated
by using either free boundary conditions or fluctuating-twist boundary
conditions,\cite{saslow:92,olsson:95} where location-dependent twists
are applied along the boundaries. The former have the problem that
corrections to scaling are huge for numerically accessible system
sizes and the latter generally overestimate the stiffness of the
system. 

Our results with periodic boundary conditions show that the
mean of the helicity modulus does not have critical scaling, thus
preventing us from determining the location of the glass transition. In
addition, strong fluctuations in the mean develop below the transition
temperature.  In contrast, the typical (median) value shows the correct
scaling, i.e., the data cross at the known value of the transition
temperature. Furthermore, we demonstrate that the data are well
fitted by extreme-value distributions, commonly used in economics
studies,\cite{gumbel:58,gumbel:60,rmetrics} explaining the strong
fluctuations in the mean as well as providing detailed estimates
of the critical parameters via a finite-size scaling analysis of
the location parameter of the helicity modulus distributions (most
probable value).

The gauge glass has attracted interest in the context
of disordered high-temperature superconductors. The phase
diagram of such disordered type-II materials exhibits numerous
phases\cite{blatter:94,brandt:95,nattermann:00,blatter:03} 
(see Fig.~\ref{fig:pd}) and, despite intensive research, several features
of the different phase diagrams of high-temperature superconductors
in a magnetic field remain to be fully understood.  In the presence
of disorder the Abrikosov lattice ceases to be perfect.  For low
enough temperatures and fields (above $H_{{\rm c}1}$), a Bragg
glass emerges\cite{blatter:03} characterized by the appearance of
algebraic Bragg peaks when spectroscopic measurements are performed,
even though the flux line lattice is disordered. These are lost when
either the disorder or the temperature is increased. Increasing
the temperature yields a vortex liquid, where flux lines fluctuate subject
to a random disorder potential. Increasing the disorder yields a vortex
glass phase\cite{fisher:89a,giamarchi:94} where the order of the flux
line lattice is destroyed and the vortex lines are strongly pinned
to the impurities with no spatial order -- similar to standard spin
glasses\cite{binder:86} (see Fig.~\ref{fig:pd}).  Signatures of a
vortex glass phase have been observed experimentally by studying the
$I$--$V$ characteristics of YBCO samples. There the vanishing of the
resistivity for fields larger than $H_{{\rm c}1}$ has been interpreted
as the onset of a vortex glass phase.\cite{koch:89,gammel:91,olsson:91}
Furthermore, a recent microscopic study\cite{divakar:04} on LSCO
samples using muon-spin rotation experiments and small-angle neutron
scattering shows the transition from a Bragg to a vortex glass phase.

Fisher {\em et al.}\cite{fisher:91b} argue that the superfluid
density $\rho_{\rm s}$ is finite in the thermodynamic limit,
with finite-frequency corrections that only vanish as an inverse
power of $\ln(\omega)$. In contrast, Korshunov\cite{korshunov:01}
recently predicted that the superfluid density tends to {\em zero}
logarithmically in $\omega$.  Because the superfluid density can be
directly related to the helicity modulus, it is tempting to describe
the strongly disordered vortex glass phase by numerically studying the
gauge glass model, although certain aforementioned provisions need
to be considered: the gauge glass model does not include a field
anisotropy and it does not have transverse screening. We use our
results for the helicity modulus in the gauge glass to discuss the
superfluid density in the vortex glass, and thus, test if the superfluid
density vanishes in the thermodynamic limit or if it stays finite
when the system is in {\em equilibrium} ($\omega = 0$). Our results
from equilibrium Monte Carlo simulations of the gauge glass for the
helicity modulus show that for the system sizes studied the superfluid
density remains finite in the thermodynamic limit.

The paper is structured as follows: In Sec.~\ref{sec:model} we
introduce the model, observables, and the numerical method used. In
Sec.~\ref{sec:results} we present our results, followed by concluding
remarks in Sec.~\ref{sec:conclusions}.

\section{Model, Observables, and Numerical Details}
\label{sec:model}

The Hamiltonian of the gauge glass is given by
\begin{equation}
{\mathcal H} = -\sum_{\langle i,j\rangle} J_{ij}\cos(\phi_i - \phi_j - A_{ij}),
\label{hamiltonian}
\end{equation}
where the sum ranges over nearest neighbors on a cubic lattice in
three space dimensions of size $N = L^3$, and $\phi_i$ represent the angles
of the $XY$ spins. For the gauge glass, $J_{ij} = J$ $\forall$ $i$, $j$ 
in Eq.~(\ref{hamiltonian}) and in this work we set $J = 1$.
Periodic boundary conditions (PBCs) are applied. 
The $A_{ij}$ represents the line integral of the vector potential
between sites $i$ and $j$, and we chose the $A_{ij}$ from a uniform
distribution in $[0,2\pi]$ but with the constraint that $A_{ij} =
- A_{ji}$.\cite{katzgraber:01a} In three space dimensions, the model
exhibits a finite-temperature spin-glass transition\cite{olson:00}
at $T_{\rm c} \approx 0.47 \pm 0.03$. Note that when screening is
added to the Hamiltonian in Eq.~(\ref{hamiltonian}), the transition
to a glass phase is lost.\cite{wengel:96}

The helicity modulus $Y$ is the second-order derivative of the free
energy $F$ with respect to an infinitesimal twist $\Theta$ to the
boundaries,\cite{reger:91} i.e., $Y = \partial^2 F / \partial \Theta^2
|_{\Theta \rightarrow 0}$, 
\begin{equation} 
\begin{split}
Y = \frac{1}{L^2}
\Bigl[
\sum_i \langle \cos(\Delta_i) \rangle  - 
\frac{1}{T} & \sum_{i,j} \langle \sin(\Delta_i) \sin(\Delta_j) \rangle  \\
& - \langle \sin(\Delta_i) \rangle \langle \sin(\Delta_j) \rangle
\Bigr].
\label{eq:helicity}
\end{split}
\end{equation}
In Eq.~(\ref{eq:helicity}), $\langle \cdots \rangle$ represents a
thermal average and $\Delta_i = \phi_i - \phi_{i+x} - A_{ii+x}$,
where the twist is performed along the $x$ axis.\cite{dirs} Because a
disorder average of the free energy $F$ is independent of the twist
angle, we expect that
\begin{equation}
[Y]_{\rm av} = 0 \;\;\;\;\;\;({\rm PBC}),
\label{eq:yave}
\end{equation}
where $[\cdots]_{\rm av}$ represents a disorder average.
This is not necessarily the case for the {\em typical} (median)
value $[Y]_{\rm typ}$.  If we study the distribution $P(Y)$ of $Y$,
then we expect a large positive contribution due to spin waves. In
addition, abrupt changes in the slope of the free energy $F$ due to
sudden vortex rearrangements in certain disorder configurations will
generate rare, very negative values of the helicity modulus $Y$, thus
generating a long negative tail in the distribution\cite{reger:91}
(see Fig.~\ref{fig:histo_0.409}).  This immediately poses the
question of whether the different moments of the distribution are
properly defined.  Especially for low temperatures $T \ll T_{\rm c}$,
we expect the tail of the distribution to be most pronounced due to
the low probability of vortex rearrangements.

In order to better quantify the effects of the tail of the
distribution, we empirically fit the data to a generalized
extreme-value distribution,\cite{gumbel:58,gumbel:60} which,
in general, is the limiting distribution of the maxima of a
sequence of independent and identically distributed random
variables. Because of this property, generalized extreme-value
distributions are used as an approximation to model the maxima
of long (finite) sequences of random variables that can be,
for example, found in the analysis of stock market data or time
series in flat-histogram methods.\cite{dayal:04,alder:04} Note that
Bertin and Clusel\cite{bertin:06} have shown that sums of correlated
variables also yield limiting distributions, which can be described by
extreme-value distributions. Therefore an underlying extreme process
is not necessary to obtain an extreme-value limiting distribution.
In addition to the aforementioned applications, extreme-value
distributions have recently been used to study different problems in
other areas of science ranging from physics and astronomy to biology
applications. In the present context, the use of extreme-value
distributions is motivated by the extremely skewed shape of the
helicity modulus distribution as well as the strong fluctuations
found in the mean and the standard deviation.

The cumulative (integrated) generalized extreme-value distribution is 
given by
\begin{equation}
H_{\rm C}(x) = \exp\left[-\left( 
1 + \xi \frac{x - \mu}{\beta}
\right)^{-1/\xi}\right] \; ,
\label{eq:evd}
\end{equation}
where $\mu$ is the ``location parameter'' (most probable value),
$\beta$ the ``scale parameter'' (standard deviation of the peak),
and $\xi$ is the ``shape parameter'' which indicates the asymptotic
behavior of the tail of the distribution. The generalized extreme-value 
distribution is then given by
\begin{equation}
H(x) = \frac{dH_{\rm C}(x)}{dx},
\label{eq:evdd}
\end{equation}
where $H_{\rm C}(x)$ is given by Eq.~(\ref{eq:evd}). In Eq.~(\ref{eq:evdd}),
when $\xi < 0$ we obtain a thin-tailed (decay faster than exponential)
Weibull distribution. If $\xi = 0$, we obtain a Gumbel extreme-value
distribution (exponential tail), whereas if $\xi > 0$, we obtain a
fat-tailed Fr\'echet distribution where the tail falls off slower
than an exponential. This has the effect that when $\xi > 0$, the
$m$th moment of the Fr\'echet distribution exists only if $\xi <
1/m$; i.e., if $\xi > 1/2$, the variance of the distribution is not
defined, and if $\xi > 1$ the mean is not properly defined either.
Here, we expect that the distribution of the helicity modulus is well
fitted by an extreme-value distribution, i.e.,
\begin{equation}
P(Y) = H(-Y) \; ,
\label{eq:evd2}
\end{equation}
where $H(x)$ is given by Eq.~(\ref{eq:evdd}).

Finally, the superfluid density $\rho_{\rm s}$ is related to the
helicity modulus via\cite{reger:91}
\begin{equation}
\rho_{\rm s} = L^{2 - d} Y,
\label{eq:sfden}
\end{equation}
where $d$ represents the space dimension. In the present case, $d =
3$ and thus $\rho_{\rm s} = L^{-1}Y$.  Therefore, by investigating
the properties of the helicity modulus of the gauge glass, we can also
obtain information about the superfluid density of the model.

The simulations are performed using the parallel tempering
Monte Carlo method.\cite{hukushima:96,marinari:98b} To
ensure that the data are in equilibrium, we perform a logarithmic
binning of the moments of different observables (such as the
currents\cite{reger:91,olson:00,katzgraber:02a} and the helicity
modulus defined above) and ensure that the data are independent of
Monte Carlo steps for the last three bins.\cite{equil} In addition, we
ensure that the acceptance probabilities for the parallel tempering
moves are always greater than $0.3$. In order to speed up the
simulations, we discretize the angles of the $XY$ spins to $N_{\phi} =
512$ values, a number large enough to avoid crossover effects to other
models.\cite{cieplak:92} Finally, to ensure that the local spin flips
are accepted often enough at low temperatures ($\sim 20$\%), we select
the proposed new angle for a spin from a temperature-dependent window
around the original angle, i.e., $\Delta \phi \propto T$ with a minimum
angle for low temperatures of $5^\circ$. The simulation parameters
are presented in Table \ref{tab:simparams}. Note that to
avoid statistical bias in the computation of the helicity modulus,
{\em four} replicas of the system have been used for each temperature 
simulated.

The data are fitted to a generalized extreme-value distribution
using the statistics package R (see Refs.~\onlinecite{rmetrics} and
\onlinecite{r}) using the \texttt{fgev} optimization routine in the
\texttt{evd} package.

\begin{table}
\caption{
Parameters of the simulation. $N_{\rm sw}$ is the total number of
Monte Carlo sweeps, $T_{\rm min}$ is the lowest temperature simulated,
and $N_T$ is the number of temperatures used in the parallel tempering
method (Ref.~\onlinecite{temps}) for each system size $L$. The highest
temperature simulated is $T_{\rm max} = 0.947 \gg T_{\rm c}$. For
each system size, $N_{\rm sa} = 10^4$ samples have been simulated in
order to probe the tails of the distribution in detail.
\label{tab:simparams}
}
\begin{tabular*}{\columnwidth}{@{\extracolsep{\fill}} c l l r }
\hline
\hline
$L$  & $N_{\rm sw}$      & $T_{\rm min}$ & $N_{T}$\\
\hline
$ 3$ & $6.0 \times 10^3$ & $0.050$       & $53$   \\
$ 4$ & $2.0 \times 10^4$ & $0.050$       & $53$   \\
$ 5$ & $6.0 \times 10^4$ & $0.050$       & $53$   \\
$ 6$ & $2.0 \times 10^5$ & $0.050$       & $53$   \\
$ 8$ & $1.2 \times 10^6$ & $0.050$       & $53$   \\
$12$ & $1.2 \times 10^6$ & $0.303$       & $38$   \\
\hline
\hline
\end{tabular*}
\end{table}

\section{Results}
\label{sec:results}

\begin{figure}[!tbp]
\includegraphics[width=\columnwidth]{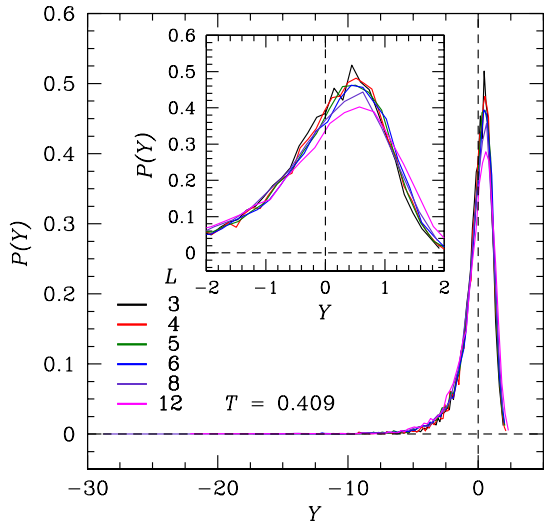}
\vspace*{-1.0cm}

\caption{(Color online)
Distribution of the helicity modulus $P(Y)$ for different system
sizes $L$ at $T = 0.409 < T_{\rm c}$. The data show a pronounced peak
close to $Y \gtrsim 0$ (see inset) and a long tail for $Y < 0$ which
grows with increasing system size, thus suggesting that the data are
extreme-value distributed. The dashed lines are guides for the eyes.
}
\label{fig:histo_0.409}
\end{figure}

\begin{figure}[!tbp]
\includegraphics[width=\columnwidth]{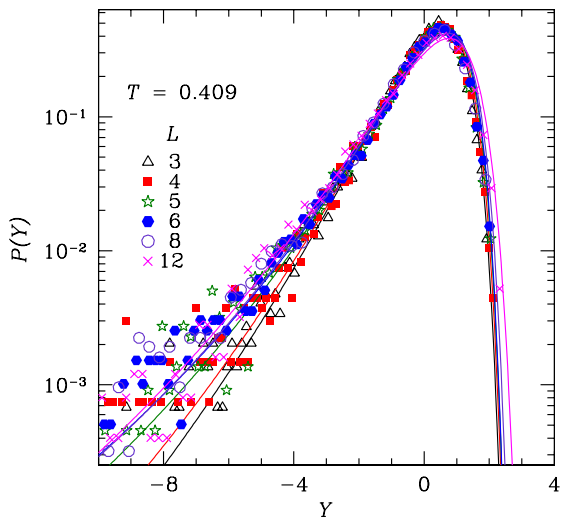}
\vspace*{-1.0cm}

\caption{(Color online)
Data for the distribution of the helicity modulus $P(Y)$ shown in
Fig.~\ref{fig:histo_0.409} in a linear-logarithmic plot. The curvature
in the tails indicates that the data are Fr\'echet distributed with a
positive shape parameter $\xi$ for $T = 0.409 < T_{\rm c}$. The width
of the tail increases slightly with increasing system size $L$. The
error bars are not shown for clarity.
}
\label{fig:loghisto_0.409}
\end{figure}

\begin{figure}[!tbp]
\includegraphics[width=\columnwidth]{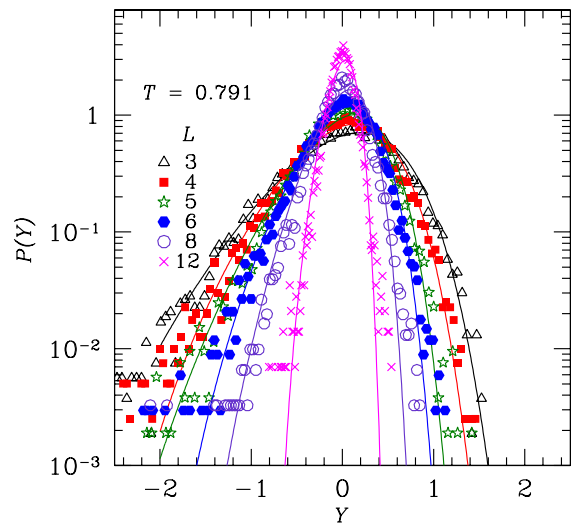}
\vspace*{-1.0cm}

\caption{(Color online)
Linear-logarithmic plot of the distribution of the helicity modulus $P(Y)$ for
$T = 0.791 > T_{\rm c}$. The tails of the distribution decay faster
than exponential, thus suggesting that the data are well described
by a thin-tailed Weibull distribution with $\xi < 0$. The error bars
are not shown for clarity.
}
\label{fig:histo_0.791}
\end{figure}

Figure \ref{fig:histo_0.409} shows the distribution of the helicity
modulus in the disorder for different system sizes at $T = 0.409$,
which is below the known value of $T_{\rm c} = 0.47(3)$.\cite{olson:00}
The data show a peak for $Y \gtrsim 0$ and a long tail that
extends to large negative values, in agreement with the discussion in
Sec.~\ref{sec:model}. The behavior of the tail can be studied in more
detail in a semilogarithmic plot, see Fig.~\ref{fig:loghisto_0.409}.
For $T < T_{\rm c}$, the tail decays slower than exponential, thus
illustrating the possibility of moments of the distribution not being
properly defined. In contrast, for $T > T_{\rm c}$, the tail of the
distribution decays faster than exponential, as can be seen, for
example, for $T = 0.791$ in Fig.~\ref{fig:histo_0.791}.

\begin{figure}[!tbp]
\includegraphics[width=\columnwidth]{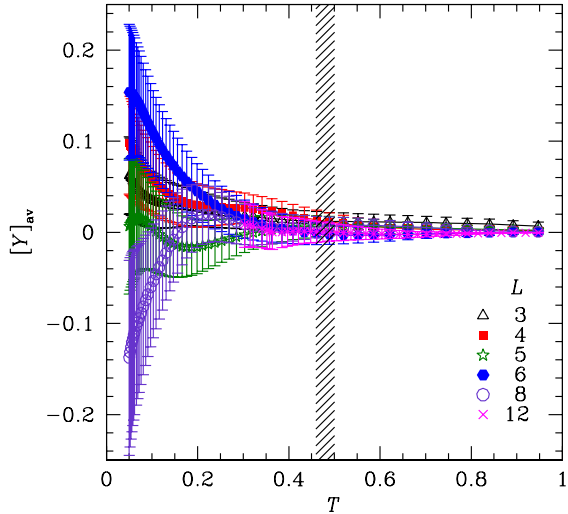}
\vspace*{-1.0cm}

\caption{(Color online)
Average helicity modulus $[Y]_{\rm ave}$ as a function of temperature
for several system sizes. The average should be zero within
error bars for all temperatures. The data show that this is the case 
for $T > T_{\rm c}$, whereas for $T < T_{\rm c}$, due to the fat tails of
the distributions (see Fig.~\ref{fig:histo_0.409}), the fluctuations in
the mean as well as the standard deviation are extremely large. Note
that the data show no signature of the phase transition at $T_{\rm c}
\approx 0.47$ (shaded area).
}
\label{fig:ave_helicity}
\end{figure}

\begin{figure}[!tbp]
\includegraphics[width=\columnwidth]{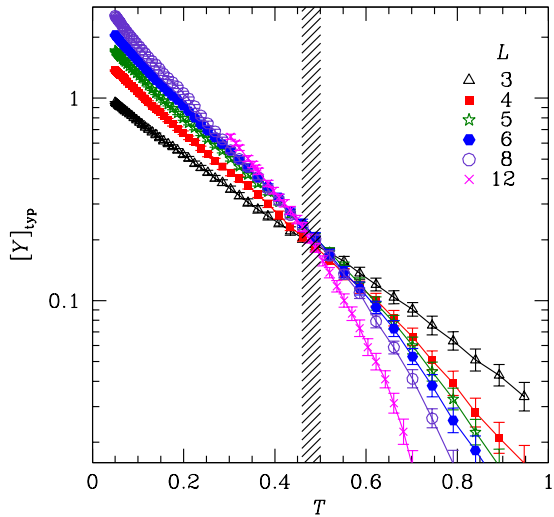}
\vspace*{-1.0cm}

\caption{(Color online)
Typical (median) helicity modulus $[Y]_{\rm typ}$ as a function of
temperature for different system sizes. The data are clearly nonzero
in equilibrium and display the correct finite-size scaling; i.e.,
the data cross at $T_{\rm c} \approx 0.45$  -- $0.50$ (shaded area).
}
\label{fig:typical_hel}
\end{figure}

In Fig.~\ref{fig:ave_helicity}, we show the average helicity modulus
as a function of temperature for different system sizes. Following
previous arguments, we expect $[Y]_{\rm av} = 0$. This is the case for
high enough temperatures, yet for low temperatures the data fluctuate
strongly with increasing system size due to the thick tails of the
distributions, even for a large number of disorder realizations.
Because we expect $[Y]_{\rm av} = 0$, the data show no evidence of
the phase transition; i.e., there is no crossing of the data for
different system sizes as one would expect from a finite-size scaling
analysis. This is not the case for the typical value (median) of the
distribution (Fig.~\ref{fig:typical_hel}), where the data are clearly
nonzero and cross approximately at the known value of the transition
temperature.

In order to improve on these results and properly account for the long
tails in the distribution of the helicity modulus, we empirically fit
the data for the distributions at different temperatures and system
sizes to a generalized extreme-value distribution [Eq.~(\ref{eq:evd})];
these are the solid lines in Figs.~\ref{fig:loghisto_0.409} and
\ref{fig:histo_0.791}. The data are well represented by the fit,
although in the tails the fitting routine slightly underestimates
the tail at extremely low temperatures; however, this does not bias the
results because the location parameter is mainly determined by the
peak in the distribution.  

\begin{figure}[!tbp]
\includegraphics[width=\columnwidth]{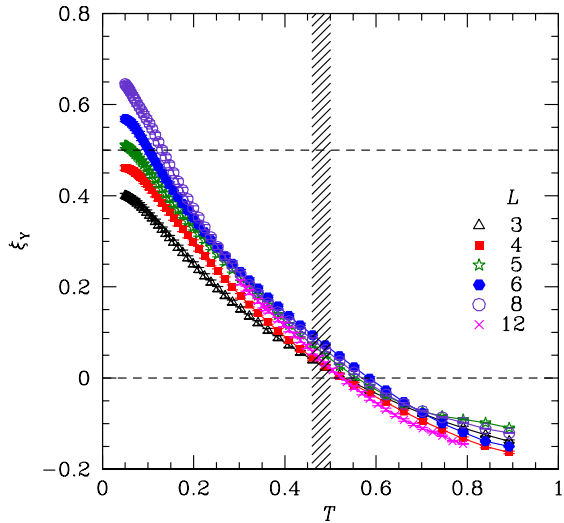}
\vspace*{-1.0cm}
                                                                                
\caption{(Color online)
Shape parameter $\xi_{\rm Y}$ of the helicity modulus distribution
as a function of temperature for different system sizes. Note
that $\xi_{\rm Y} < 0$ for $T \gtrsim T_{\rm c}$ (shaded region); i.e.,
the distribution is thin-tailed and the moments of the distribution
are properly defined. For $T \lesssim T_{\rm c}$, the shape parameter
is positive; i.e., the distribution is fat tailed. Already
for $T \approx 0.10$ and $L \gtrsim 8$, the standard deviation of the
distribution is not properly defined ($\xi_{\rm Y} > 1/2$, marked by the 
upper horizontal dashed line).  This explains the strong fluctuations 
observed in the average helicity presented in Fig.~\ref{fig:ave_helicity}.
}
\label{fig:shape_hel}
\end{figure}

We first study the shape of the helicity
modulus distributions via the shape parameter $\xi_{\rm Y}$. In
Fig.~\ref{fig:shape_hel}, we show the shape parameter as a function
of temperature for the different system sizes simulated. For $T
\gtrsim T_{\rm c}$, the shape parameter is negative for all $L$, thus
indicating that when we are above the transition temperature the data
are well described via a thin-tailed Weibull distribution. This is
not the case for $T \lesssim T_{\rm c}$, where the shape parameter
increases for all system sizes when the temperature decreases to
zero. In particular, already for small system sizes $L \approx 8$
and $T \approx 0.10$, the shape parameter exceeds $1/2$; i.e.,
the variance of the distribution is ill-defined.  This explains
the diverging fluctuations of the average helicity modulus in
Fig.~\ref{fig:ave_helicity} for $T \rightarrow 0$. By extrapolating
the data, we expect that for $T \approx 0.10$ and $L \gtrsim 15$,
the mean will not be properly defined any longer (since, then, $\xi
> 1$).  

From a physical standpoint, a thin-tailed distribution for
$T > T_{\rm c}$ suggests that vortex jumps, while they might occur,
tend to be exponentially suppressed. This is not the case for $T <
T_{\rm c}$, where the tail of the distribution remains ``fat;'' i.e.,
arbitrarily large vortex rearrangements can occur.  Note that for $T
\rightarrow \infty$, we expect the distribution in the thermodynamic
limit to become Gaussian, which corresponds to $\xi = -1/2$.

\begin{figure}[!tbp]
\includegraphics[width=\columnwidth]{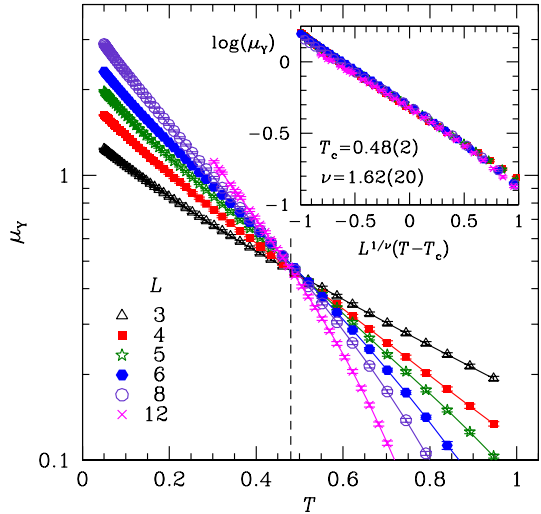}
\vspace*{-1.0cm}

\caption{(Color online)
Location parameter of the helicity modulus distribution $\mu_{\rm Y}$
as a function of temperature for different system sizes. The
data are nonzero and cross cleanly at $T_{\rm c} = 0.48(2)$. The
inset shows a finite-size scaling analysis of the data according
to Eq.~(\ref{eq:scale}). $L = 3$ has been omitted in the finite-size
scaling analysis due to strong
corrections to scaling. The observed scaling of the data suggests that
generalized extreme-value distributions describe the helicity
modulus distribution well. The vertical dashed line in the main panel 
at $T = 0.48$ is a guide for the eyes.
}
\label{fig:location_hel}
\end{figure}

Because in an extreme-value distribution the location parameter $\mu$
represents the most probable value (similar to the median), we show
in Fig.~\ref{fig:location_hel} the location parameter of the helicity
modulus distribution, $\mu_{\rm Y}$, as a function of temperature for
different system sizes.  The data intersect cleanly at $T = 0.48(2)$,
thus showing clear evidence of the glass transition. The fact that
the data intersect so cleanly suggests that the helicity modulus
distribution can be well described by extreme-value distributions. In
the inset of Fig.~\ref{fig:location_hel}, we show a finite-size scaling
plot of the data according to
\begin{equation}
\mu_{Y} \sim \tilde{M}[L^{1/\nu}(T - T_{\rm c})] ,
\label{eq:scale}
\end{equation}
where $\tilde{M}$ is a scaling function and $\nu$ is the critical
exponent of the correlation length.\cite{yeomans:92} The best
scaling collapse is determined by a nonlinear minimization
routine\cite{katzgraber:06} using the software package
R.\cite{r,rmetrics} The data scale well and we estimate $T_{\rm
c} = 0.48(2)$ and $\nu = 1.62(20)$, in agreement with previous
estimates of the critical exponents by Olson and Young computed
from a finite-size scaling of the currents [$T_{\rm c} = 0.47(3)$ and
$\nu = 1.39(20)$, see Ref.~\onlinecite{olson:00}].\cite{twodim} Note
that a recently introduced extended scaling scheme\cite{campbell:06}
might yield slightly different estimates for the critical exponents;
however, we expect the location of the transition to remain unchanged.

\begin{figure}[!tbp]
\includegraphics[width=\columnwidth]{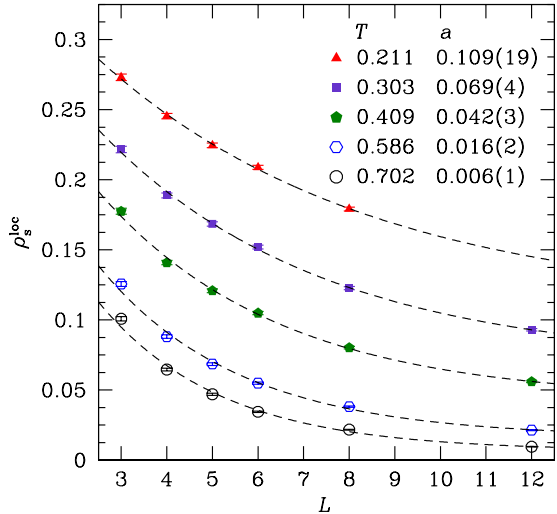}
\vspace*{-1.0cm}

\caption{(Color online)
Superfluid density $\rho_{\rm s}^{\rm loc}$ computed from the location
parameter $\mu_{\rm Y}$ as a function of system size for different
temperatures.  The dashed lines are fits according to $\rho_{\rm
s}^{\rm loc}(L) = a + b \exp(-cL)$.
}
\label{fig:rho_s_location}
\end{figure}

Since the average helicity modulus is zero and the location parameter
shows the correct critical behavior, we compute the superfluid density
$\rho_{\rm s}^{\rm loc}$ from Eq.~(\ref{eq:sfden}) by replacing $Y$
with $\mu_{\rm Y}$ and plot it as a function of system size $L$ for
different temperatures in Fig.~\ref{fig:rho_s_location}. The data
decay with increasing system size for all temperatures below and
above the critical point and are well fitted by a phenomenological
constant$+$exponential behavior, i.e., $\rho_{\rm s}^{\rm loc}(L)
= a + b \exp(-cL)$ (dashed lines in Fig.~\ref{fig:rho_s_location})
with fitting probabilities\cite{press:95} $Q \sim 0.3$ for $T
\rightarrow 0$.\cite{fit} Therefore our results suggest that
$\rho_{\rm s}^{\rm loc}(L = \infty)$ is nonzero for the system
sizes studied; for reference, Fisher {\em et al.}~predict
$\rho_{\rm s}^{\rm loc} \sim a + b/\log(\omega/\omega_0)^x$ where
$\omega_0$ is a characteristic frequency and $x$ an exponent,
and Korshunov\cite{korshunov:01} predicts $a = 0$.  We emphasize
that our results do not rule out the scenario as predicted by
Korshunov,\cite{korshunov:01} where the superfluid density is zero
in equilibrium: The work of Korshunov is based on the existence of a
hierarchical distribution of metastable states, and it is unclear if
such an approximation would be valid for the gauge glass. In addition,
large system sizes, which are numerically inaccessible, are required
to probe a logarithmic behavior in detail.

The advantage of fitting the data with extreme-value distributions
is also illustrated when studying the superfluid density: While
the data for the typical estimate also display qualitatively the
same behavior as the data computed from the location parameter,
the fluctuations are considerably larger for the former (not shown)
and fitting the data to extract any limiting behavior is impossible.

\section{Conclusions}
\label{sec:conclusions}

We show that the distribution of the helicity modulus in the three-dimensional
Gauge glass is well described by an extreme-value distribution 
and demonstrate that the typical value of the helicity modulus is 
nonzero, exhibiting critical scaling, whereas the behavior of the average 
value is an artifact of the boundary conditions: the average estimate is 
zero and displays strong fluctuations below the transition temperature. These
fluctuations can be explained by fitting the data to an extreme-value
distribution and by studying the shape parameter of the distribution.
Below $T_{\rm c}$ the data are fat tailed, and thus, different moments
of the distribution are ill defined.

In addition, we have computed the superfluid density from the location
parameter of the extreme-value distribution and shown that it is finite
in the thermodynamic limit.  In this work, the equilibrium properties of
the gauge glass [$\omega \rightarrow 0$ limit before $L \rightarrow
\infty$ limit] have been studied.  Another question of interest
is if the limits $L \rightarrow \infty$ and $\omega \rightarrow 0$
are interchangeable.  Assuming the gauge glass to be a good effective
model of the vortex glass phase in high-temperature superconductors,
our results for the gauge glass imply that the superfluid density
of the vortex glass is finite in the thermodynamic limit when the
system is in equilibrium. Much larger system sizes are required
to answer this question beyond reasonable doubt.

Finally, it might be of interest to perform similar studies on
improved models of vortex glasses, such as the recently introduced
model from Ref.~\onlinecite{kawamura:00} which includes the disorder
via the couplings $J_{ij}$ between the $XY$ spins 
[see Eq.~(\ref{hamiltonian})]. Furthermore, the
effects of screening on the gauge glass should be revisited at very
low temperatures.  We hope that this work will find application to
other studies of observable distributions in physics where fat tails
are involved.

\begin{acknowledgments}

We would like to thank I.~A.~Campbell, V.~B.~Geshkenbein, H.~Kawamura,
S.~E.~Korshunov, M.~K\"orner, Y.~Ozeki, S.~Trebst, M.~Troyer,
M.~Wallin, G.~T.~Zim\'anyi, and, in particular ,A.~P.~Young for helpful
discussions. H.G.K.~thanks Heron Island for their hospitality during
the last stages of the paper. The simulations were performed on the
Asgard cluster at ETH Z\"urich. This work has been supported in part
by the Swiss National Science Foundation.

\end{acknowledgments}

\bibliography{comments,refs}

\begin{thebibliography}{45}
\bibitem[{ani()}]{aniso}
\bibinfo{note}{Note that high-temperature superconductors also possess an
  intrinsic material anisotropy which can be rescaled away
  (Ref.~\onlinecite{blatter:94}).}
\bibitem[{dir()}]{dirs}
\bibinfo{note}{A twist along the $y$ and $z$ directions would yield the same
  results.}
\bibitem[{equ()}]{equil}
\bibinfo{note}{For $L \le 8$ we use the same simulation parameters as in
  Ref.~\onlinecite{katzgraber:01a} where a thorough equilibration test was
  performed.}
\bibitem[{two()}]{twodim}
\bibinfo{note}{In two space dimensions the gauge glass does not have a
  transition at finite temperatures (see Refs.~\onlinecite{nishimori:93} and
  \onlinecite{katzgraber:03a}). We have also computed the helicity modulus
  distribution of the two-dimensional gauge glass. Unpublished data for the
  location parameter of the helicity modulus distribution supports a
  zero-temperature transition.}
\bibitem[{fit()}]{fit}
\bibinfo{note}{We have also tried other fitting functions inspired by the
  results of Ref.~\onlinecite{fisher:89a}, i.e., $\rho_{\rm s}^{\rm loc} \sim a
  + [\tilde{b}/\log(L/L_0)^x]$ where $L_0$ is a characteristic length scale and
  $x$ an exponent. In the previous expression we have assumed a power-law
  relationship between the driving frequency $\omega$ and the system size $L$,
  i.e., $\omega \sim L^{-z}$. Unfortunately, attempting to fit the numerical
  data to the aforementioned functional form yielded extremely poor fitting
  probabilities ($Q \sim 10^{-6}$).}
\end{thebibliography}

\end{document}